\documentstyle[12pt]{article}
\begin{document}
\newcommand\be{\begin{equation}}
\newcommand\ee{\end{equation}}
\voffset-2.5cm
\hoffset-1.5cm
\baselineskip 24pt

\title{Molecule Formation and  the Farey Tree in the
One-Dimensional Falicov-Kimball Model}
\author{\sl C. Gruber,\ \sl D. Ueltschi \\
Institut de Physique Th\'{e}orique, \\
Ecole Polytechnique F\'{e}d\'{e}rale de Lausanne, \\
PHB-Ecublens, CH-1015 Lausanne, Switzerland
\and
\sl J. J\c{e}drzejewski \\
Institute of Theoretical Physics,\\
University of Wroc{\l}aw, \\
Plac Maxa  Borna 9, 50--204 Wroc{\l}aw, Poland \\
}
\date{}
\maketitle
\noindent
{Key words: Falicov--Kimball model, ground states,  phase diagram, \\
phase separations, molecules}
\\
\\
\begin{abstract}
The ground state configurations of the one--dimensional
Falicov--Kimball model are studied exactly with numerical calculations
revealing unexpected effects for small interaction
strength. In neutral systems we observe molecular formation,
phase separation and changes in the conducting properties; while
in non--neutral systems the phase diagram exhibits Farey tree
order (Aubry sequence) and a devil's staircase structure.
Conjectures are presented for the boundary of the segregated domain
and the general structure of the ground states.
\end{abstract}

\newpage

\section{Introduction}

The study of the general properties of fermion systems, such as
metal-insulator transitions, crystal formation and transitions to
mixed-valence states, on the basis of microscopic models, is a domain
of vigorous research. Among the models considered nowadays growing
attention is being paid to a simple model proposed almost a quarter of
a century ago by Falicov and Kimball \cite{falkim}. This model was put
forward to describe metal-insulator transitions in transition-metal
and rare-earth materials, where an analysis of the experimental data
suggested that these transitions are of purely electronic origin.

The first investigations of the Falicov-Kimball model, carried out by
means of approximate methods, like mean field or coherent phase
approximations \cite{khom}, led to contradictory results and reduced
the general interest in the model. However in 1986 Kennedy and Lieb
\cite{kenlieb} and independently Brandt and  Schmidt \cite{brasch}
obtained rigorous results showing the existence of a phase transition
at sufficiently low temperatures. They  considered the simplest
version of the model with only one band and spinless fermions, later
called the spinless Falicov-Kimball model. Even this simplified
version has numerous, physically sound, interpretations \cite{khom},
\cite{kenlieb}.
In one of them, related to crystal formation, the system is viewed as
consisting of quantum electrons hopping over a lattice and
classical ions occupying some of the lattice sites.
The interesting case occurs when
the lattice consists of two equivalent sublattices and electrons are
allowed to hop only from a site belonging to one sublattice to
a nearest neighbour that belongs to the other sublattice.
The only interaction is the on-site attraction (or repulsion)
between  electrons and ions and the hard core repulsion between ions.

If the densities are equal to $1/2$,
or equivalently, if the chemical potentials  correspond
to the hole-particle symmetry point (half filled band case),
then  it has been proved that in dimensions two and higher
this simple interaction leads to an order-disorder transition as the
temperature is varied \cite{kenlieb}, \cite{brasch} .

In the low temperature phase the ions arrange themselves into a
periodic structure, where they occupy only one of the sublattices.
An important point to achieve this remarkable result is
the fact that, in distinction to other lattice fermion models,
a study of the Falicov-Kimball model can be reduced to a study of
tight-binding Schr\"{o}dinger equations in a variety of potentials.
At each site these potentials assume only two values, say $0$ or
$-U$, that correspond to the absence or presence of an ion at
this site. Let us underline that contrary to band
theory, where the potential in the Schr\"{o}dinger equation is fixed,
the potential is here a variable and the problem is to find that
potential whose contribution to the partition function is dominant (the so
called annealed problem).

One consequence of this fact is that the model can be studied by
comparing the ground state energies corresponding to certain classes
of potentials (ion configurations), i.e. by constructing
restricted phase diagrams for a finite or infinite system.
Initially the method of restricted phase diagrams was applied to
a $2$-dimensional system in \cite{jll} and to a 1-dimensional system
in \cite{frefal}.

In recent years numerous papers devoted to the spinless
Falicov-Kimball model appeared. They dealt mainly with the following
topics:--various generalizations of the first rigorous results
(quoted above) \cite{brasch2},\cite{gijl},\cite{lebmac},\cite{mm},
--investigations of the segregation principle
(first formulated by Freericks and Falicov \cite{frefal}) \cite{gru},
\cite{lem}, \cite{bra},
--investigations of the effective interaction between the ions
\cite{barsub}, \cite{gjl}, \cite{glm}, \cite{ken},
--attempts to find out the general structure of the ground state phase
diagram of the one-dimensional system, in particular new candidates
for low-temperature configurations of the ions \cite{frefal},
\cite{llj}.
We shall refer to the papers quoted here in due course.

This paper belongs to the last of the mentioned groups.
It is organized as follows. In section 2 we provide the reader with
the relevant informations, definitions and notations. Then in
section 3 we describe the results obtained in the canonical ensemble;
after that in section 4 the results obtained in the grand canonical
ensemble. Finally in section 5 we summarize our observations,
conclusions and conjectures.

\section{Definitions and notations}

The object of this work is to study the ground state properties of
the one--dimensional spinless Falicov--Kimball model. This model
describes a system of itinerant spinless quantum fermions interacting
with classical ions on a one--dimensional lattice $\Lambda$.
Whenever the lattice has a finite number $|\Lambda|$ of lattice sites
we impose periodic boundary conditions. The Hamiltonian is

\begin{equation}
\label{ham}
H_{\Lambda}=\displaystyle{ \sum\limits_{x\in \Lambda} \left(
a^{*}_{x} a_{x+1} + a^{*}_{x+1} a_{x} - U w(x) a^{*}_{x} a_{x}
\right),}
\end{equation}
where $a^{*}_{x}, a_{x}$ are the creation and annihilation operators
of itinerant spinless electrons and $w(x)=0$ or $1$ is the number
of ions at the lattice site $x$.
The total number of electrons
$N_{e}=\sum \limits_{x\in \Lambda} a^{*}_{x} a_{x}$ and
the total number of ions (in the configuration $w$)
$N_{i}(w)=\sum \limits_{x \in \Lambda} w(x)$ are the conserved
quantities. The function $w(x), x\in \Lambda$, is called the ion
configuration.
The electrons can hop only between nearest neighbour
sites and the corresponding kinetic energy matrix elements are set
to unity. Therefore in $H_{\Lambda}$ there is only one independent
parameter expressed in these units: the electron-ion interaction $U$.

For a given configuration $w(x)$ the
Hamiltonian (\ref{ham}) is the second quantized form of the
one-particle operator whose matrix elements are

\begin{equation}
\label{matrix_h}
h_{xy}=t_{xy} - U w(x) \delta_{xy},
\end{equation}
where
\begin{equation}
\label{matrix_t}
\begin{array}{l}
t_{xy} =
\cases{ 1 &\hbox{if} $y = x\pm 1$ \cr
0 &\hbox{otherwise}. }
\end{array}
\end{equation}
In the following, when we speak of the spectrum of the configuration
$w$ we mean the spectrum of this one-particle Hamiltonian.
The bands (gaps) in this spectrum are referred to as the bands (gaps)
of the configuration $w$.
The ground state energy $E^{U}(w; N_{e})$ of the system (\ref{ham})
corresponding to $N_{e}$ electrons and the ion configuration $w$
is equal to the sum of the $N_{e}$ lowest energy levels of the
single--particle Hamiltonian (\ref{matrix_h}).

Applying to (\ref{ham}) the unitary hole--particle transformation
with respect to the ions we obtain

\begin{equation}
\label{symm_i}
E^{(U)}\left(w^{*};N_{e}\right)
= E^{(-U)}\left( w;N_{e}\right) - UN_{e},
\end{equation}
where $w^{*}(x)=1-w(x)$,
while with respect to the electrons we get

\begin{equation}
\label{symm_e}
E^{(U)}\left(w;N_{e}\right)
= E^{(-U)}\left( w; |\Lambda| - N_{e}\right) - UN_{i}(w).
\end{equation}
Therefore we can reduce the range of $N_{e}$ and $N_{i}$ considered
and arbitrarily fix the sign of $U$ (in the sequel we
drop the superscript $U$ by $E$).
We choose $U>0$, i.e. the particles attract each other at the
same  lattice site.

We study also the infinite systems in the thermodynamic limit.
The limit is taken in such a way that the particle densities per
site $N_{e}/| \Lambda |$, $N_{i}/| \Lambda |$ and the ground
state energy density $E(w;N_{e})/| \Lambda |$ tend to the
electron density $\rho_{e}$, the ion density $\rho_{i}$ and the
ground state energy density $e(w;\rho_{e})$.
For the periodic ion configuration $w$ the Green function of the
tight binding Schr\"{o}dinger equation and consequently the density of
states $n(w)$, $\rho_{e}$ and $e(w;\rho_{e})$ can be determined
exactly \cite{lyz}. However the corresponding formulae contain the
band edges of the spectrum of $w$, which are zeroes of the polynomial
whose order coincides with the period of $w$. At this point it is
necessary to perform numerical calculations.

The problem we want to investigate is to find the \underline{ground
state configurations} (g.s.c.), i.e. the zero temperature phase
diagram.
In the canonical formalism one is given the electron and ion
densities $(\rho_{e},\rho_{i})$ and the task is to find the
configurations $\bar{w}$ that minimize the energy $E(w, N_{e})$,
$N_{i}(w)=N_{i}$, or the energy density $e(w, \rho_{e})$,
$\rho_{i}(w)=\rho_{i}$. In the following sections we shall construct
"restricted phase diagrams", which amounts to minimizing the set of
functions $e(w,\rho_{e})$, where $w$ runs over some specified class
of admissible configurations, for example the class of all periodic
configurations with period smaller than some fixed value.
Since the minimum of a set of convex functions is not necessarily
convex, one has to take the convex envelope of the minimum. For those
values $\rho_{e}$ where the convex envelope does not coincide with the
minimum of the $e(w, \rho_{e})$, the g.s.c. is then a mixture of two
or more configurations. The configuration $w$ of ions in the mixture
$w_{1}\& w_{2}$ is defined as follows: The finite lattice $\Lambda$
is partitioned into two parts $\Lambda_{1}$ and $\Lambda_{2}$;
the restriction of $w$ to $\Lambda_{1}$ and $\Lambda_{2}$ is $w_{1}$
and $w_{2}$ with $|\Lambda_{1}|/|\Lambda|=\alpha$
and

\begin{equation}
\label{rhoi_mix}
\rho_{i} = \alpha \  \rho_{i}(w_{1})
+ (1 - \alpha)
\ \rho_{i}(w_{2}).
\end{equation}
The thermodynamic limit is then taken  keeping $\alpha$ fixed.
In particular the mixture of the full and empty configurations
is the so called {\em segregated configuration}\cite{frefal}
where all the ions clump together.
The mixture $w\& -$ will play an important role
in our studies of neutral systems (section 3).
The density of states of the mixture $w\& -$ is

\begin{equation}
\label{den_mix}
n_{\alpha} (\mu) = \alpha \ n (w; \mu )
+ ( 1 - \alpha ) n (- ; \mu ) \ ,
\end{equation}
where
\begin{equation}
\label{den-}
n(- ; \mu ) = \pi^{-1}\  Re \left( 4 - \mu^{2} \right)^{-1/2} .
\end{equation}
and $n(w;\mu)$ is given in \cite{lyz}.
Consequently the electron and ion densities are:

\begin{equation}
\label{rhos_mix}
\rho_{e} = \alpha \  \rho_{e}(w)
+ (1 - \alpha)
\ \rho_{e} (-), \qquad
\rho_{i} = \alpha \ p/q \ ,
\end{equation}
where
\begin{equation}
\label{rho_w}
\rho_{e}(w) = \rho_{e}(w; \mu_{F}) = \int\limits^{\mu_{F}}
_{- \infty} \ n(w; \mu) d\mu
\end{equation}
and $\rho_{e}(-)$ is given by eq.(\ref{rho-}). Similarly the energy
density of this mixture is

\begin{equation}
\label{en_mix}
e = \alpha \  e(w) + (1 - \alpha ) \ e(-) \ ,
\end{equation}
where
\begin{equation}
\label{en_w}
e(w) = e(w; \mu_{F}) =
\int\limits^{\mu_{F}}_{- \infty}
\ \mu \ n(w; \mu) d\mu
\end{equation}
and $e(-)$ is given by eq.(\ref{en-}).

In the grand canonical formalism one is given the electron and ion
chemical potentials $(\mu_{e}, \mu_{i})$ and the task is to find the
configurations $\bar{w}$ that minimize the free energy density

\begin{equation}
\label{fen_w}
f\left( w;\mu_{e}, \mu_{i} \right)
=
e\left( w ; \mu_{e} \right) -
\mu_{e} \ \rho_{e}
\left( w; \mu_{e} \right)
- \mu_{i} \ \rho_{i}(w) \ .
\end{equation}
Again we shall construct "restricted phase diagrams" which amounts to
minimizing $f(w; \mu_{e}, \mu_{i})$ over some class of configurations.
In this case there is no need to take
the convex envelope since the free energy density (\ref{fen_w}) is a
concave function of the chemical potentials.

In the following sections we shall consider special classes of
configurations. It is useful to define them here and to introduce
an appropriate notation. There are two translationally invariant
configurations: the {\em full configuration}, denoted by $+$,
where all the sites of the lattice are occupied ($\rho_{i}=1$) and
the {\em empty configuration}, denoted by $-$, with no ions
($\rho_{i}=0$).
The unique band of the empty configuration extends from $-2$ to $2$
(while the band of the full configuration is translated by $-U$).
At the Fermi level $\mu_{F}$ the electron density is

\begin{equation}
\label{rho-}
\rho_{e}\left(-;\mu_{F}\right)
= \
\left\{
\begin{array}{lll}
0 \qquad &\hbox{if}
\qquad
& \mu_{F} \leq -2,
\cr\cr
\pi^{-1} \ arc \cos \left(-\mu_{F}/2\right)
&\hbox{if}
& |\mu_{F}|\leq 2,
\cr\cr
1 &\hbox{if}
&\mu_{F} \geq 2.
\end{array}
\right.
\end{equation}
and the corresponding ground state energy density is

\begin{equation}
\label{en-}
e\left( - ; \mu_{F} \right)
= \
\left\{
\begin{array}{lll}
-\pi^{-1} \sqrt{4-\mu^{2}_{F} }\qquad
&\hbox{if}
\qquad
&|\mu_{F} | \leq 2,
\cr\cr
0 && \hbox{otherwise.}
\end{array}
\right.
\end{equation}

A periodic ion configuration that does not belong to one of the
two classes considered, with period $q$ and $p$ ions per
unit cell ($\rho_{i}=p/q$) is denoted by $p/q$.

The unit cell of the
{\em crenel configuration} $\{ p/q \}$ consists of $p$ consecutive
sites occupied by the ions while the remaining $q-p$ sites are empty.
The {\em atomic most homogeneous configuration} with $\rho_{i}=p/q$
is denoted by $[p/q]$. In this configuration the ions in the unit
cell are distributed in such a manner that the distances between two
consecutive ions are either $d$ or $d+1$ with $d\leq q/p < d+1$.
Furthermore the distribution of the distances $d$ and $d+1$ has to
be most homogeneous  \cite{lem}, \cite{hub}. More precisely, with $p$
relatively prime to $q$, the
position of the ions in the unit cell is given by
$k_{j}$ solutions of the equations
$p\, k_{j}=j \; mod \: q$, $j=0, 1, \ldots, p-1$.

Similarly we define the
{\em n--molecule most homogeneous configuration} $[p/q]_{n}$,
$n\geq 2$, where $\rho_{i}=p/q$ and $p$ is a multiple of $n$.
This configuration is defined in the same way as $[p/q]$, but
replacing "ion" by "$n$--molecule" that consists of $n$ consecutive
ions. More precisely, with $p/n$ relatively prime to $q$,
the position of the ions in the unit cell is given by
$k_{j}$ solutions of the equations
$(p/n) k_{j}=j \; mod \: q$, $j=0, 1, \ldots, p-1$.
In particular $[n/q]_{n}$ is identical with $\{n/q\}$.
In the following we denote the $n$--molecules by $H_{n}$.

The following example illustrates the above definitions: for $p=6$
and $q=10$ the unit cell for the crenel configuration $\{6/10\}$ is
$[\bullet\bullet\bullet\bullet\bullet\bullet\circ\circ\circ\circ]$,
for the atomic most homogeneous configuration $[6/10]$ it is
$[\bullet\bullet\circ\bullet\circ\bullet\bullet\circ\bullet\circ]$,
i.e. it is the extension of the unit cell corresponding to $[3/5]$,
for the $2$--molecule most homogeneous configuration $[6/10]_{2}$ it
is $[\bullet\bullet\circ\circ\bullet\bullet\circ\bullet\bullet\circ]$,
for the $3$--molecule most homogeneous configuration $[6/10]_{3}$ it
is
$[\bullet\bullet\bullet\circ\circ\bullet\bullet\bullet\circ\circ]$.
Finally as an example of a periodic configuration $6/10$ that is not
a member of the above mentioned classes we can take the one with the
unit cell
$[\bullet\bullet\circ\bullet\bullet\circ\bullet\circ\bullet\circ]$.
The list of all the periodic ion configurations that enter into our
phase diagrams is given in Table \ref{a}.
The mixture of the configuration $w=p/q$ and the empty configuration
is denoted $p/q\& -$ or $p/q\& vacuum$ (if the Fermi level $\mu_{F}$ is
smaller than $-2$).

\begin{table}
\scriptsize
\begin{tabular}{cc}\cr
{
\begin{tabular}{|c|c|l|}\hline\hline
&&\cr
$\rho_{e}$  &$\rho_{i}$ & unit cell \cr
&&\cr
\hline\hline
 1/11
 & $[2/11]_{2}$
 &
\cr
&&\cr

 & $[3/11]_{3}$
 &
\cr
&&\cr
   1/10
 &$[2/10]_{2}$
 &
\cr
&&\cr

 & $[3/10]_{3}$
 &
\cr
&&\cr

 & $[4/10]_{4}$
 &
\cr
&&\cr

 & $[5/10]_{5}$
 &
\cr
&&\cr
  1/9
 & $[2/9]_{2}$
 &
\cr
&&\cr
 & $[3/9]_{3}$
 &
\cr
&&\cr

 & $[4/9]_{4}$
 &
\cr
&&\cr
 1/8
 & $[2/8]_{2}$
 &
\cr
&&\cr
 & $[3/8]_{3}$
 &
\cr
&&\cr
 & $[4/8]_{4}$
 &
\cr
&&\cr
 1/7
 & $[2/7]_{2}$
 &
\cr
&&\cr
 & $[3/7]_{3}$
 &
\cr
&&\cr
 & $[4/7]_{4}$
 &
\cr
&&\cr
 1/6
 & $[2/6]_{2}$
 &
\cr
&&\cr

& $[3/6]_{3}$
 &
\cr
&&\cr
 2/11
 & $[4/11]_{2}$
 & $\bullet\bullet\circ\circ\circ\bullet\bullet\circ\circ\circ\circ$
\cr
&&\cr
  1/5
 & $[1/5]$
 &
\cr
&&\cr
\hline
\end{tabular}
}
&
{
\begin{tabular}{|c|c|l|}\hline\hline
&&\cr
$\rho_{e}$  &$\rho_{i}$ & unit cell \cr
&&\cr
\hline\hline
   1/5
 & $[2/5]_{2}$
 &
\cr
&&\cr

 & $[3/5]_{3}$
 &
\cr
&&\cr
  4/19
 & $[4/19]$
 &
\cr
&&\cr
  3/14
 & $[3/14]$
 &
\cr
&&\cr
  2/9
 & $[2/9]$
 &$\bullet\circ\circ\circ\circ\bullet\circ\circ\circ$
\cr
&&\cr

 & $3/9$
 & $\bullet\bullet\circ\circ\circ\bullet\circ\circ\circ$
\cr
&&\cr

 & $[4/9]_{2}$
 & $\bullet\bullet\circ\circ\bullet\bullet\circ\circ\circ$
\cr
&&\cr

 & $5/9$
 &$\bullet\bullet\circ\circ\bullet\bullet\bullet\circ\circ$
\cr
&&\cr

 & $[6/9]_{3}$
 &$\bullet\bullet\bullet\circ\bullet\bullet\bullet\circ\circ$
\cr
&&\cr
   1/4
 & $[1/4]$
 &
\cr
&&\cr

 & $[2/4]_{2}$
 &
\cr
&&\cr
   3/11
 & $[3/11]$
 & $\bullet\circ\circ\circ\bullet\circ\circ\circ\bullet\circ\circ$
\cr
&&\cr

 & 4/11
 & $\bullet\bullet\circ\circ\bullet\circ\circ\circ\bullet\circ\circ$
\cr
&&\cr

 & 5/11
 & $\bullet\bullet\circ\circ\bullet\bullet\circ\circ\bullet\circ
    \circ$
\cr
&&\cr

 & $[6/11]_{2}$
 & $\bullet\bullet\circ\circ\bullet\bullet\circ\circ\bullet\bullet
    \circ$
\cr
&&\cr

 & 7/11
 &$\bullet\bullet\bullet\circ\bullet\bullet\circ\circ\bullet\bullet
   \circ$
\cr
&&\cr
   2/7
&$[2/7]$
& $\bullet\circ\circ\circ\bullet\circ\circ$
\cr
&&\cr

& 3/7
& $\bullet\bullet\circ\circ\bullet\circ\circ$
\cr
&&\cr

&$[4/7]_{2}$
&$\bullet\bullet\circ\circ\bullet\bullet\circ$
\cr
&&\cr
\hline
\end{tabular}
}
\end{tabular}
\caption{(a) The periodic ion configurations that appear in the
             restricted phase diagrams.}
\end{table}

\setcounter{table}{0}
\begin{table}
\scriptsize
\begin{tabular}{cc}\cr
{
\begin{tabular}{|c|c|l|}\hline\hline
&&\cr
$\rho_{e}$  &$\rho_{i}$ & unit cell \cr
&&\cr
\hline\hline
  2/7
&$[5/7]$
&$ \bullet\bullet\bullet\circ\bullet\bullet\circ$
\cr
&&\cr
 3/10
&$[3/10]$
&$\bullet\circ\circ\circ\bullet\circ\circ\bullet\circ\circ$
\cr
&&\cr

& $4/10$
& $ \bullet\bullet\circ\circ\bullet\circ\circ\bullet\circ\circ$
\cr
&&\cr

&$5/10$
&$\bullet\bullet\circ\circ\bullet\bullet\circ\bullet\circ\circ$
\cr
&&\cr

 &$[6/10]_{2}$
 &$\bullet\bullet\circ\circ\bullet\bullet\circ\bullet\bullet\circ$
 \cr
 &&\cr
  1/3
 &$[1/3]$
 &
 \cr
&&\cr

 &$[2/3]_{2}$
 &
 \cr
&&\cr
  3/8
 & $2/8$
 &$\bullet\circ\circ\bullet\circ\circ\circ\circ$
 \cr
&&\cr

 &$[3/8]$
 &$\bullet\circ\circ\bullet\circ\circ\bullet\circ$
 \cr
&&\cr

 &$4/8$
 &$\bullet\bullet\circ\bullet\circ\circ\bullet\circ$
 \cr
&&\cr

 &$5/8$
 &$\bullet\bullet\circ\bullet\bullet\circ\bullet\circ$
 \cr
&&\cr

 &$[6/8]_{2}$
 &$\bullet\bullet\circ\bullet\bullet\circ\bullet\bullet$
 \cr
&&\cr
  2/5
 &$1/5$
 &$\bullet\circ\circ\circ\circ$
 \cr
&&\cr

 &$[2/5]$
 &$\bullet\circ\circ\bullet\circ$
 \cr
&&\cr

 &$3/5$
 &$\bullet\bullet\circ\bullet\circ$
 \cr
&&\cr

 &$[4/5]_{2}$
 &$\bullet\bullet\circ\bullet\bullet$
 \cr
&&\cr
 5/12
 & $[5/12]$
 &$\bullet\circ\circ\bullet\circ\circ\bullet\circ\bullet\circ\bullet
   \circ$
\cr
&&\cr

 & $6/12$
 &$\bullet\circ\circ\bullet\circ\bullet\bullet\circ\bullet\circ
   \bullet\circ$
\cr
&&\cr

 & $[7/12]$
 &$\bullet\bullet\circ\bullet\circ\bullet\bullet\circ\bullet\circ
   \bullet\circ$
\cr
&&\cr
\hline
\end{tabular}
}
&
{
\begin{tabular}{|c|c|l|}\hline\hline
&&\cr
$\rho_{e}$  &$\rho_{i}$ & unit cell \cr
&&\cr
\hline\hline
 5/12
 & $8/12$
 &$\bullet\bullet\bullet\bullet\circ\bullet\circ\bullet\bullet\circ
   \bullet\circ$
\cr
&&\cr

 & $9/12$
 & $\bullet\bullet\bullet\bullet\circ\bullet\bullet\bullet\bullet
    \circ
    \bullet\circ$
\cr
&&\cr
  3/7
 &$2/7$
 &$\bullet\circ\bullet\circ\circ\circ\circ$
 \cr
&&\cr

 &$[3/7]$
 &$\bullet\circ\bullet\circ\bullet\circ\circ$
 \cr
&&\cr

 &$4/7$
 &$\bullet\circ\bullet\circ\bullet\bullet\circ$
 \cr
&&\cr

 &$5/7$
 &$\bullet\circ\bullet\circ\bullet\bullet\bullet$
 \cr
&&\cr
 4/9
 &$3/9$
 &$\bullet\circ\bullet\circ\bullet\circ\circ\circ\circ$
 \cr
&&\cr

 &$[4/9]$
 &$\bullet\circ\bullet\circ\bullet\circ\bullet\circ\circ$
 \cr
&&\cr

&$5/9$
&$\bullet\circ\bullet\circ\bullet\circ\bullet\bullet\circ$
\cr
&&\cr

&$6/9$
&$\bullet\circ\bullet\circ\bullet\bullet\bullet\bullet\circ$
\cr
&&\cr
 5/11
 & $3/11$
 & $\bullet\circ\bullet\circ\bullet\circ\circ\circ\circ\circ\circ$
\cr
&&\cr

 & $4/11$
 & $\bullet\circ\bullet\circ\bullet\circ\bullet\circ\circ\circ
    \circ$
\cr
&&\cr

 & $[5/11]$
 & $\bullet\circ\bullet\circ\bullet\circ\bullet\circ\bullet\circ
    \circ$
\cr
&&\cr

 & $[6/11]$
 & $\bullet\circ\bullet\circ\bullet\circ\bullet\circ\bullet
    \circ\bullet$
\cr
&&\cr
 & $7/11$
 & $\bullet\bullet\bullet\circ\bullet\circ\bullet\circ\bullet
    \circ\bullet$
\cr
&&\cr

 & $8/11$
 & $\bullet\bullet\bullet\bullet\bullet\circ\bullet\circ\bullet
    \circ\bullet$
\cr
&&\cr
1/2
&$[1/2]$
&
\cr
&&\cr
&&\cr
&&\cr
&&\cr
&&\cr
\hline
\end{tabular}
}
\end{tabular}
\caption{(b) The periodic ion configurations that appear in the
             restricted phase diagrams; continuation.\label{a}}
\end{table}

\newpage

\section{Neutral systems. Phase diagrams in the canonical ensemble}

In this section we study the ground state properties of the neutral
systems in the canonical ensemble. If the system is finite we thus
take $N_{e}=N_{i}=N$ and we are interested in the configurations
$\bar{w}$ that minimize the energy $E(w;N)$.

If the system is infinite, then $\rho_{e}=\rho_{i}=\rho$ and
the problem is to find the configurations $\bar{w}$ that
minimize the energy density $e(w;p/q)$. We shall first consider only
periodic configurations with $\rho=p/q$, $p$ and $q$ being relatively
prime integers.

For large $U$ and rational densities it was first
conjectured \cite{frefal} and then rigorously proved that the g.s.c.
is the most homogeneous configuration of atoms \cite{lem}.
Since this result was established using the perturbation series of
$h(w)$ in powers of $U^{-1}$, the question naturally arises whether
the restriction to large $U$ is technical or fundamental. Until now
it has been usually conjectured that such a theorem should be valid
for all values of $U$  \cite{lem}, \cite{lieb}.
Indeed, the results of
Kennedy and Lieb \cite{kenlieb} and Brandt and Schmidt \cite{brasch}
for the half--filled band case $(\rho=1/2)$ and the numerical studies
of Freericks and Falicov \cite{frefal} for $\rho_{i}=1/2$, $1/3$
seemed to support this conjecture.

However the analysis of a finite number of ions and electrons on an
infinite lattice (with $N_{e}=N_{i}$), while providing a qualitative
explanation of this result, suggests also that it cannot
remain valid for small $U$. Indeed, for large $U$
and to leading order in $U^{-1}$ the energy of the neutral system
is given by  an effective two--body potential of the form
$2(d+1)\exp [-\lambda (2d+1)]$, where $d$ is the distance between the
ions \cite{glm}. In other words, for large $U$ an equal number of ions
and electrons will form neutral atoms which repel each
other with an effective two--body potential that is convex and
decreasing.
Using the result of Hubbard \cite{hub} we thus conclude that for large
$U$ the g.s.c. of the neutral system with a finite density should be
the most homogeneous configuration.

On the other hand, one can easily check that for $U\leq 2/\sqrt{3}$
the energy of the neutral $2$--molecule is smaller than the
energy of two neutral atoms that are infinitely separated.
Indeed the ground state energy $E^{atom}$ of the neutral atom is

\be
\label{en_at}
E^{atom} = - \sqrt{4 + U^{2}},
\ee
while the energy $E^{2-molecule}$ of the neutral $2$--molecule is

\be
\label{en_2m}
E^{2-molecule} = \cases{\displaystyle{ - 2 {U^{3}\over U^{2} - 1}}
&if \quad $U \geq$ 2, \cr\cr
- \displaystyle{\left( U + 2\right)^{2} \over U + 1}
&if \quad  $U \leq 2.$
}
\ee
\\
Moreover for $U\leq2$ the $2$--molecule has only one bound state.

More data are contained in Fig. 1, where the energy
of two ions and two electrons on the infinite lattice is shown
as a function of the distance between the ions for different values
of $U$. For large values of $U$ the ground state consists of two
neutral atoms  at infinite distance. On the other hand, for
small values of $U$ this happens when
one $2$--molecule with one bound electron and one diffusion
electron is formed.
Similarly, Fig. 2 shows that the ion configuration that minimizes the
energy of three ions and three electrons depends on the value of $U$.
There are three kinds of ground states:\\
-- three neutral atoms at infinite distance ($U$ large),
-- one $2$--molecule with one bound state and one neutral atom at
infinite distance,
-- one $3$--molecule with one bound state ($U$ small).
\begin{figure}
\vspace{9truecm}
\caption{Energy per ion for two ions and two electrons versus
the distance between the ions.}
\end{figure}
\begin{figure}
\vspace{9truecm}
\caption{Energy per ion for three ions and three electrons
versus the two distances between the ions.}
\end{figure}

Finally we can see in Fig. 3 that the energy between two
$2$--molecules (and 4 electrons) presents the same properties as
the energy between two ions: for sufficiently large $U$ it is convex
and decreasing and thus the ground state consists of two
$2$--molecules infinitely separated, while for $U$ sufficiently
small the ground state consists of one $4$--molecule with one bound
state. Although we have not checked it, we expect that the energy
between two $n$--molecules (and $2n$ electrons) has similar
properties.
\begin{figure}
\vspace{9truecm}
\caption{Energy per ion for two $2$--molecules and four
electrons versus the distance between the molecules.}
\end{figure}
To conclude the above considerations of neutral systems with
a finite number of ions,
let us find the $n$--molecule with the lowest energy
per ion as a function of $U$. The result is shown in Table \ref{b}.
\begin{table}
\begin{center}
\begin{tabular}{|l|c|}\hline\hline
$U$--interval &Type of $n$--molecule \cr\hline\hline
&\cr
$(1.154,\, \; \infty \; \;)$  & $H_{1}$  \cr
&\cr
$(0.511,\, 1.154)$   & $H_{2}$  \cr
&\cr
$(0.279,\, 0.511)$   & $H_{3}$  \cr
&\cr
$(0.175,\, 0.279)$   & $H_{4}$  \cr
&\cr
$(0.119,\, 0.175)$   & $H_{5}$  \cr
&\cr
$(0.086,\, 0.119)$   & $H_{6}$  \cr
&\cr
$(0.065,\, 0.086)$   & $H_{7}$  \cr
&\cr
$(0.051,\, 0.065)$   & $H_{8}$  \cr
&\cr
$(0.041,\, 0.051)$   & $H_{9}$  \cr
&\cr
\hline\hline
\end{tabular}
\caption{$n$--molecules with the smallest energy per ion versus $U$.
 \label{b}}
\end{center}
\end{table}
We note that for each  $n$--molecule there is an interval of $U$
where its energy is minimal. Moreover in this interval the
$n$--molecule has only one bound state.
We then conjecture that for given $U$, the neutral system with zero
density of particles will form $n$--molecules, with $n=n(U)$,
which are infinitely separated.

The above results suggest that, for sufficiently small $U$,
the neutral systems with \underline{nonzero density}
might realize their ground
state by forming $n$--molecules, $n>1$, that are most homogeneously
distributed.
To test this idea we have considered a finite system of 512 sites
and studied the restricted phase diagram for configurations of the
form $[p/q]_{n}$. For example, for $\rho_{i}=9/128$ ($N_{i}=N_{e}=36$)
we have considered the ion configurations whose unit cells
are displayed in Table \ref{c}.
\begin{table}
\begin{center}
\begin{tabular}{|c|l|}\hline
\hline
&\cr
type of molecule &\hfil {unit cell}\hfil \cr
&\cr
\hline\hline
&\cr
$H_{1}$
   & $(\bullet)(13 \circ)(\bullet)(13\circ)
         (\bullet)(13\circ)(\bullet)(14\circ)(\bullet)
         (13\circ)(\bullet)(13\circ) $\cr
   & $(\bullet)(13\circ)(\bullet)(13\circ)(\bullet)(14\circ) $\cr
	 &\cr
	 $H_{2}$
         & $(2\bullet)
         (26\circ)
         (2\bullet)
                (27\circ)
         (2\bullet)
         (26\circ)
         (2\bullet)
         (27\circ)
         (2\bullet)
         (26\circ)$
\cr
& $ (2\bullet)
         (27\circ)
        (2\bullet)
         (26\circ)
           (2\bullet)
         (27\circ)
          (2\bullet)
         (26\circ)$
	 \cr
	 &\cr
$H_{3}$
&  $(3\bullet)
(39\circ)
         (3\bullet)
(40\circ)
         (3\bullet)
(40\circ)$
\cr
	 &\cr
$H_{4}$
& $
(4\bullet)
(52\circ)
(4\bullet)
(53\circ)
(4\bullet)
(53\circ)
(4\bullet)
(53\circ)
(4\bullet)
(53\circ)$
\cr
& $(4\bullet)
(53\circ)
(4\bullet)
(53\circ)
(4\bullet)
(53\circ)
(4\bullet)
(53\circ)$
\cr
&\cr
\hline
\end{tabular}
\caption{Periodic configurations  for $\rho=9/128$ taken into account
in Fig. 4. In our notation, for instance $(3\bullet)$ stands for three
consecutive sites occupied by ions while $(3\circ)$ stands for three
consecutive empty sites. \label{c}}
\end{center}
\end{table}
\noindent
The results are given in Fig. 4. To determine the finite size effect,
the transition line between the atomic configurations $H_{1}$ and
the $2$--molecule configurations $H_{2}$ has been computed for a
system of 200 sites and no significant difference was observed.
We also compared the energy of the finite system with the energy of
the corresponding periodic configuration of the infinite system
(e.g. see Table \ref{f} and Table \ref{g}).
{}From this analysis we can thus conclude that the usual
conjecture concerning the ground state of a neutral system (mentioned
at the beginning of this section) cannot be correct:
{\em for sufficiently small $U$ and sufficiently small densities the
most homogeneous configuration of atoms cannot be
the ground state configuration}.
\begin{figure}
\vspace{7truecm}
\caption{Restricted phase diagram with respect to
$n$--molecule most homogeneous configurations, $n\leq 4$,
over the lattice of $512$ sites.
The points are special values discussed in the text.}
\end{figure}

Of course from this analysis we cannot conclude that the
molecular configurations $[p/q]_{n}$ are the true g.s.c.
Therefore for some special values of $U$ and $\rho$ (Fig. 4),
and for an \underline{infinite system},
we have computed the energies of \underline{all} the periodic
configurations with period smaller or equal to some $Q$.
The configurations $\overline{w_{p}}$
with the smallest energy and their energy are given in Table \ref{e}.
\begin{table}
\begin{center}
\begin{tabular}{|l|r@{/}l|c|l|r@{.}l|c|}\hline\hline
U & \multicolumn{2}{c|}{$ \rho $}
& {period $\leq Q$ :} & \hfil {unit cell}\hfil
& \multicolumn{2}{c|}{energy} & {energy ($ [ 1/p ] $) } \cr\hline\hline
0.10 & 1 & 5 & 20 & {$\bullet \circ\circ\circ\circ $}
& -- 0 & 378 \ 561  &   \cr
0.10 & 1 & 6 & 24 &{$\bullet\bullet\circ\circ\circ\circ
\bullet\bullet(16 \circ)$}
& -- 0 & 321 \ 413  & -- 0.321 \ 386   \cr
0.10 & 1 & 7 & 21 &{$\bullet\bullet\bullet (18\circ)$}
& -- 0 & 278 \ 539  & -- 0.278 \ 511  \cr
0.20 & 1 & 6 & 24 &{$\bullet\bullet \circ\circ\circ\circ
\bullet\bullet(16\circ) $}
& -- 0 & 325 \ 073 & -- 0.324 \ 972  \cr
0.20 & 1 & 7 & 21 &{$\bullet\bullet \circ
\circ\circ\circ\circ\bullet(13\circ) $}
& -- 0 & 281 \ 342  & -- 0.281 \ 238 \cr
0.50 &1 &5 &20 &{$\bullet \circ\circ\circ\circ$}
& -- 0 & 401 \ 926 \cr
0.50 &1 &6 &24
&$\bullet\bullet\circ\circ\circ\bullet\bullet
(17\circ)$
& -- 0 & 338 \ 876  & --0.338 \ 655 \cr
0.50  &1 & 7  &21
&{$\bullet\bullet
\circ\circ\circ\circ\circ
\bullet
(13\circ)$}
& -- 0 & 292 \ 371  & --0.291 \ 988 \cr
0.54 &1 &6 &18
&$\bullet\bullet\circ\circ\circ\circ\bullet
(11\circ)$
& -- 0 & 340 \ 847  & --0.340 \ 815  \cr
0.60 &1 &6 &18
&$\bullet\circ\circ\circ\circ\circ$
&-- 0 & 344 \ 206 &
\cr
1.00 &1 &6 &24 &  $\bullet\circ\circ\circ\circ\circ$
&-- 0 &371 \ 472  & \cr
1.00 &1 &7 &21 & $\bullet\circ\circ\circ\circ\circ\circ$
&-- 0 & 318 \ 390 &
\cr
\hline
\end{tabular}
\caption{The canonical restricted phase diagram for an infinite chain
and special values of $(U, \rho)$. The ground state configurations were
selected among all the periodic configurations whose period does not
exceed $Q$. \label{e}}
\end{center}
\end{table}

Clearly, for sufficiently small $U$ and $\rho$ the
most homogeneous configurations of atoms or molecules are not the
g.s.c.
These results show that theorem 1 of Freericks and Falicov
\cite{frefal} is not always correct: for $\rho=p/q$ there exist
periodic configurations whose energy is smaller than the energy
of atomic configurations and whose period is larger than  $q$
(similar results hold true for smaller values of $U$).
However, we shall see in the sequel that, if the true g.s.c. is
periodic then it is given by the construction of Freericks and Falicov
\cite{frefal} and this remains true  in the non--neutral case
($\rho_{e}\neq \rho_{i}$).

Furthermore Table \ref{e} shows that for
small densities and small $U$ the particles tend to arrange
themselves into molecules (as it was  the case at zero density)
and lead us to think that the g.s.c. could be mixtures of periodic
configurations of molecules and the empty configuration.
To check this guess we have considered a system of 6 ions and 6
electrons on a lattice of 42 sites, i.e. $\rho=1/7$.
The energies of all ion configurations have been computed and the
g.s.c. selected (see Table \ref{f}).
\begin{table}
\begin{center}
\begin{tabular}{|c|c|c|}\hline\hline
U &unit cell & energy \cr\hline\hline
&&\cr
 0.1  &{$\bullet\bullet(5\circ)
\bullet\bullet(5\circ)
\bullet\bullet(26\circ)$}
 & -- 0.278 \ 265
\cr
&&\cr
0.2 & $\bullet\bullet(4\circ)
\bullet\bullet(4\circ)
\bullet\bullet(28\circ)$
&-- 0.281 \ 258 \cr
&&\cr
\hline\hline
\end{tabular}
\caption{Ground state configurations of the neutral system with
42 sites and density $\rho=1/7$. \label{f} }
\end{center}
\end{table}
We see that the g.s.c. of this
finite system ($\rho=1/7$)  are the mixtures of
$\{2/7\}\&-$ or $\{2/6\}\&-$.
However Fig. 4 suggest that the mixture of
$\{3/6\}\&-$ could come into competition if the lattice were
sufficiently large. We have thus computed the energies of these
three mixtures for a lattice of 420 sites and for the infinite
lattice (see Table \ref{g}).
\begin{table}
\begin{center}
\begin{tabular}{|c|c|c|c|}\hline\hline
U & configuration & energy ($L=420$) & energy ($L=\infty$) \cr
\hline\hline
&&&\cr
0.1 & $\{ 2/7\} \ \& \ -$
&-- 0.278 580
&-- 0.278 679 \cr
&&&\cr
& $ \{ 3/6\}\ \& \ -$
&-- 0.278 538
&-- 0.278 535 \cr
&&&\cr
& $ \{ 2/6 \}\ \&\ -$
&-- 0.278 500
&-- 0.278 500
\cr
&&&\cr
& $ [ 1/7 ] $
&
&-- 0.278 511
\cr
&&&\cr\hline
&&&\cr
0.2  & $ \{ 3/6 \}\ \&\ - $
&-- 0.281 470
&-- 0.281 490 \cr
&&&\cr
&  $ \{ 2/6 \}\ \& \ -$
&-- 0.281 452
&-- 0.281 455
\cr
&&&\cr
&  $ \{ 2/7 \}\ \& \ -$
&-- 0.281 441
&-- 0.281 439
\cr
&&&\cr
&  $ [ 1/7 ]  $
&
&-- 0.281 238
\cr
&&&\cr\hline
\end{tabular}
\caption{Energy density of the mixture $\{p/q\}\&-$ for the neutral
system with density $\rho=1/7$, for a lattice of 420 sites and for an
infinite lattice, compared with the energy density of the atomic
most homogeneous configuration $[1/7]$. \label{g}}
\end{center}
\end{table}
These new and surprising results show that for sufficiently small
values of $U$ and $\rho$ the g.s.c. is indeed a mixture of some
periodic configuration with the empty configuration.
We have thus looked at
the restricted phase diagram for all configurations that are mixtures
of the empty configuration and a periodic configuration with period
$q\leq 7$ for $\rho=1/5, 1/6, 1/7$ and period $q\leq 10$ for
$\rho=1/10$. Note that, in particular, the most homogeneous
configurations are among those considered.
Table \ref{h} shows that the configurations minimizing the
energy are  mixtures of $\{p/q\}\&-$ (i.e. crenel configurations
followed by the empty configurations) or the most homogeneous
configurations of atoms.
\begin{table}
\begin{center}
\begin{tabular}{|c|c|c|c|c|c|c|c|c|c|c|c|}
\hline\hline
&&&&&&&&&&&\cr
$\rho\Bigg\backslash U $
&0.1 & 0.2& 0.3& 0.4 & 0.5 & 0.6 & 0.7 & 0.8 & 0.9 & 1.0 &1.1
\cr
&&&&&&&&&&&\cr
\hline\hline
&&&&&&&&&&&\cr
1/5
& $\{2/5\}$
& $\{2/5\}$ & $H_{1}$  & $H_{1}$ & $H_{1}$ & $H_{1}$ & $H_{1}$
& $H_{1}$ & $H_{1}$ & $H_{1}$ & $H_{1}$
\cr
&&&&&&&&&&&\cr
1/6
& $\{2/6\}$
& $\{2/6\}$
& $\{2/5\}$
& $\{2/5\}$
& $\{2/5\}$
& $\{2/5\}$
& $\{2/5\}$
& $H_{1}$
& $H_{1}$
& $H_{1}$
& $H_{1}$
\cr
&&&&&&&&&&&\cr
1/7
& $\{2/7\}$
& $\{3/6\}$
& $\{2/6\}$
& $\{2/6\}$
& $\{2/6\}$
& $\{2/6\}$
& $\{2/5\}$
& $\{2/5\}$
& $\{2/5\}$
& $H_{1}$
& $H_{1}$
\cr
&&&&&&&&&&&\cr
1/10
& $\{3/9\}$
& $\{3/8\}$
& $\{3/8\}$
& $\{2/8\}$
& $\{2/7\}$
& $\{2/7\}$
& $\{2/7\}$
& $\{2/6\}$
& $\{2/6\}$
& $\{2/6\}$
& $H_{1}$
\cr
&&&&&&&&&&&\cr
\hline
\end{tabular}
\caption{Restricted phase diagram for mixtures of the form
$p/q\&-$ (infinite lattice).\label{h}}
\end{center}
\end{table}
This observation suggests to restrict our attention to those
configurations that are either most homogeneous or mixtures of
the form $\{p/q\}\&-$. We have thus constructed the restricted
phase diagram for all configurations  that are either most
homogeneous or mixtures of $\{p/q\}\&-$ with $q\leq 10$.
This has been done in the range $U\in [0.01, 1.2],
\rho\in [0.1,0.3]$, taking steps $\Delta U=0.05$,
$ \Delta\rho=0.005$. We have also considered the values of $\rho$
(in the above range) of the form $a/b$ with $b\leq 29$.
In the domains that have not been expected to appear
(e.g. $\{3/6\}$, between $\{2/6\}$ and $\{2/7\}$) finer steps were
used:
$\Delta U=0.01, \Delta\rho=0.001$. Finally the transition lines
separating different domains of $2$--molecules have been computed
explicitly. The results are summarized in Fig. 5.
\begin{figure}
\vspace{7truecm}
\caption{Restricted phase diagram of the infinite neutral
system in the canonical ensemble. All the mixtures of the form
$\{p/q\}\& -$, with $q<10$, and the atomic most homogeneous
configurations whose period does not exceed $10$ were taken
into account. The dots correspond to the entries of Table 7.
The configurations found in the grand canonical analysis are located
at the vertical segments. The fat vertical segments represent the
mixtures $\{1/5\}\& \{2/5\}\& -$ and $\{2/5\}\& \{2/6\}\& -$.
The domains labeled 1 and 2 are $[4/9]_{4}$ and $[4/8]_{4}$,
respectively.}
\end{figure}

To achieve a better understanding of the transition we have
fixed $U=0.6$ and we have found that for $\rho\in [0.14,0.15]$
the mixture of $\{2/5\}\&\{2/6\}\&-$ has lower energy than
$\{2/5\}\&-$ or $\{2/6\}\&-$. Similarly for
$\rho\in [0.18,0.19]$ the mixture $\{1/5\}\&\{2/5\}\&-$ has
smaller energy than $\{1/5\}\&-$ or $\{2/5\}\&-$.
Therefore the transition lines should be replaced by transition
boundaries with mixtures of one or two periodic molecular
configurations and the empty configuration. Our investigations of
the grand canonical phase diagram, described in the sequel,
suggest that between two domains
$[p/q]_{n}\&-$ and $[p'/q']_{n}\&-$ there is the domain
$[(p+p')/(q+q')]_{n}\&-$ and so on. On the other hand between two
domains $[p/q]_{n}\&-$ and $[p'/q']_{n+1}\&-$ there is a mixture
$[p/q]_{n}\&[p'/q']_{n+1}\&-$.

Let us summarize our observations. For neutral systems
($\rho_{e}=\rho_{i}=\rho$) and for $U\leq 2/\sqrt{3}$ there is
a critical density $\rho_{c}(U)$ such that for $\rho > \rho_{c}(U)$
the g.s.c. is atomic most homogeneous (as for $U>2/\sqrt{3}$)
and thus the g.s.c. varies from point to point.
On the other hand for $\rho < \rho_{c}(U)$ larger and larger
molecules appear, the g.s.c. remains a mixture of the periodic
$n$--molecule configuration $[p/q]_{n}$ and the empty configuration,
over a certain range of densities.
As $\rho$ decreases the period $q$ increases discontinuously until a
value where a mixture $[p/q]_{(n+1)}\&-$ takes over.

Furthermore, if the g.s.c. is atomic most homogeneous, then the system
is an insulator, while if the g.s.c. is a mixture, the system is
a conductor. We thus have a transition from an insulator to a conductor
as $U$ decreases.

We now give an interpretation of our observations.
We have already noticed that a finite number of ions
and electrons on an infinite lattice will tend to make
$n$--molecules (with one bound state only) which repel each other.
For finite densities $\rho_{i}=\rho_{e}=\rho$ we would like to
understand  why the g.s.c. is a mixture of $\{n/q\}\&-$ and
what is the physical mechanism responsible for the period $q$.
Looking at the band structures of the periodic configurations
$\{n/q\}$ we notice that the Fermi level $\mu_{F}$
of the  mixture $\{n/q\}\&-$, corresponding to $\rho$, is in the
gap between the first two bands (Fig. 6) and thus we have

\be
\label{int1}
\rho_{i}=\rho = \alpha {n\over q}\ ,
\quad
\rho_{e} = \rho = \alpha {1\over q} + \left( 1 - \alpha \right)
\rho_{e} \left(- ; \mu_{F} \right).
\ee
Thus

\be
\label{int2}
\rho_{e}^{\{ n/q\}} = \alpha {1 \over q} = {\rho \over n} \ ,
\quad
\rho_{e} \left( - ; \mu_{F}\right)
= \rho {n -1 \over n - \rho q},
\ee
i.e. there is one  electron per molecule in a bound state associated
with the periodic ion configuration
while the remaining $\rho (n-1)/n$ electrons are in the extended
states associated with the empty configuration of the ions.
\begin{figure}
\vspace{9truecm}
\caption{Band structure of the configurations $\{2/5\}$ and
$\{2/6\}$ for $U$ between $0.1$ and $1.0$. The vertical lines show the
location of the Fermi level for the densities $\rho=1/10, 1/7$
and $1/6$.}
\end{figure}
The corresponding energy density of the system is

\be
\label{int3}
e = {\rho \over n}
\left( q \, e^{(1)} \left( \left\{ n/q\right\}\right)
- 2 \left( n - 1 \right)
{\sin y \over y}\right),
\ee
where $e^{(1)}(\{n/q\})$ is the energy density of the first band of
the configuration $\{n/q\}$ and $y=\pi\rho (n-1)/(n-\rho q)$.
For a given $\rho$ and $n$ the first term decreases as $q$ increases,
which can be seen as a consequence of the effective repulsion
between $n$--molecules due to the bound electrons.
On the other hand the second term increases as $q$ increases, which
reflects the fact that the electrons in the empty space tend to
increase the volume occupied.
Therefore the period $q$ that minimizes the energy density results
from the competition between the effective repulsive force between
the molecules and the pressure exerted by the electrons in the empty
configuration of the ions.

\section{ Phase diagrams in the grand canonical ensemble}

To obtain a better understanding of the phases at zero temperature,
and to avoid the difficulties associated with mixtures that one has
to consider in the canonical approach, we shall now investigate the
infinite system in the grand canonical formalism.
The problem is the following:
given the chemical potentials $(\mu_{e}, \mu_{i})$, what is the
configuration $\bar{w}$ that minimizes the free energy density
$f(w;\mu_{e}, \mu_{i})$.

In the following we consider only periodic configurations and we
construct the phase diagrams restricted to periodic configurations
with period smaller or equal to 10, for interaction $U$ taking
values 0.6, 0.4, 0.2.
Using the symmetry properties and the known results \cite{gjl} we
can restrict our investigation to the region
$\mu_{e}\in [-2-U, -U/2]$, $\mu_{i}\in [-U/2, 0]$.
We have thus constructed the restricted phase diagrams in the
following manner:
For fixed $\mu_{e}$, and given $w$, the free energy is a linear
function of $\mu_{i}$.
We can therefore calculate exactly the interval of $\mu_{i}$ for which
the configuration $\bar{w}$ yields the minimum of the free energy
(among all the configurations whose period does not exceed 10).
This was done taking steps $\Delta \mu_{e}=0.05$.
Near the end point $(\mu_{e}=\mu_{e}^{*})$ and near the
transition lines finer steps were taken.
The resulting phase diagrams are presented in Fig. 7.
To study these diagrams we have also computed the band edges of
all configurations that appear in Fig. 7 and we have marked
the intervals of $\mu_{e}$ in which they are g.s.c. (see Fig. 8).
\begin{figure}
\vspace{20truecm}
\caption{Restricted phase diagram of the infinite chain in the
grand canonical ensemble. All the periodic configurations whose
period does not exceed $10$ were taken into account.}
\end{figure}
\begin{figure}
\vspace{12truecm}
\caption{Bands of the g.s.c. that appear in the grand canonical
phase diagram (Fig. 7) for $U=0.6$. The continuous horizontal lines
mark the extents of the bands,
the dotted segments the $\mu_{e}$--extents of
the corresponding g.s.c.}
\end{figure}

Our analysis of the restricted phase diagrams, together with the
information on the band structure, shows that the
$(\mu_{e}, \mu_{i})$ plane is divided into three  main domains:
$D_{+}$, where the full configuration is the g.s.c.,
$D_{-}$, where the empty configuration is the g.s.c.
and $D$, where the g.s.c. is
different from the full or empty configuration.
In the range $\mu_{e}\leq -U-2$ we already know \cite{gjl} that for
$\mu_{i}>0$ the g.s.c. is the full configuration, for $\mu_{i}<0$
the g.s.c. is the empty configuration and
for $\mu_{i}=0$ all configurations have the same energy.
For $-U-2<\mu_{e}\leq \mu_{e}^{*}$, where $\mu_{e}^{*}<-2$
(see Fig. 7), the domains $D_{+}$ and $D_{-}$ are separated
by the curve

\be
\label{f+=0}
f\left( + ; \mu_{e}, \mu_{i} \right) =
f\left( - ; \mu_{e} , \mu_{i} \right) = 0,
\ee
i.e.

\be
\label{d+d-}
\begin{array}{rl}
 \pi \mu_{i}   & = - \sqrt{4 - {\tilde{\mu}}^{2}_{e} }
- \tilde{\mu}_{e} arc \  \cos \left( - \tilde{\mu}_{e}/2 \right) \ ,
\cr\cr
\tilde{\mu}_{e} & = \mu_{e} + U.
\end{array}
\ee
A point $(\mu_{e}, \mu_{i})$ on this curve corresponds
to a mixture of the full configuration with electron density

\be
\label{rho_e+}
\rho_{e} \left( + , \mu_{e} \right) = \pi^{-1}
arc \  \cos \left( - \tilde{\mu}_{e}/2 \right)
\ee
and the empty configuration with $\rho_{e}(-;\mu_{e})=0$ (vacuum).
This is the so called segregated configuration, in which all ions
clump together \cite{frefal}. Along the curve (\ref{d+d-}) we have

\be
\label{rhos_seg1}
\rho_{e}/\rho_{i} \leq \rho_{e}
\left( + ; \mu^{*}_{e} \right).
\ee
In other words for densities $\rho_{e}$ and $\rho_{i}$ satisfying
inequality (\ref{rhos_seg1}) the g.s.c. is the segregated
configuration.
The problem of determining the value of $\mu_{e}^{*}$ will be
addressed later on.

Fig. 8 shows that for each
configuration $\bar{w}$ appearing in the phase diagrams (Fig. 7)
the Fermi level is located in a gap of the corresponding spectrum.
This property has also been observed for $U=0.4$ while for $U=0.2$
it does not seem to be true since the boundary between two domains
with $\rho_{e}=p/q$ and $\rho_{i}=p'/q$, $(p'+1)/q$ is not
a straight line. However we have checked that this property is indeed
verified when we consider configurations with larger period.

Fig. 7 shows that the domain $D$ consists of connected
subdomains $D_{(\rho_{e}, \rho_{i})}$ characterized by
some periodic configuration of the ions and by a fixed density of
electrons $\rho_{e}$ (since the Fermi level is in the gap).
We note that for given densities $(\rho_{e}, \rho_{i})$, if there is
a periodic g.s.c. corresponding to this density, then it is associated
with one connected domain  $D_{(\rho_{e}, \rho_{i})}$.
This is a consequence of the fact the the free energy surface
is concave and at zero temperature it is piecewise affine.
It turns out also that the configuration of the ions can be
determined by the procedure given by Freericks and Falicov
\cite{frefal}:

Let $\rho_{e}=p/q$, with $p$ relatively prime to $q$
and $\rho_{i}=p_{i}/q$, where $p_{i}$ may not be
relatively prime to $q$. Define the integer $k_{j}$ by the relation
$p k_{j}=j\; mod\; q, j=0,1,2,\ldots$. Then the position of the ions
in the unit cell is given by $k_{j}$ with $j=0,1,2,\dots,p_{i}-1$.

The following structure of $D$ is remarkable:
the union $D_{\rho_{e}}$ of the domains  $D_{(\rho_{e}, \rho_{i})}$
corresponding to the same electron density $\rho_{e}=p/q$
constitutes a connected region. In this region the ion
density takes on the values $p'/q, (p'+1)/q, \ldots, p''/q$,
for some $p'$ and $p''>p'$ (see Fig. 9).
Moreover the unit cell of the configuration associated
with the ion density $\rho_{i}=(p+1)/q$ can be obtained by adding one
ion to the unit cell of the configuration associated
with $\rho_{i}=p/q$.
\begin{figure}
\vspace{9truecm}
\caption{Domains $D_{\rho_{e}}$ with $\rho_{e}=3/8$, $2/5$,
$3/7$ and $4/9$.}
\end{figure}

Another property of $D$ is related with the observation that the
domains $D_{n}$, where the ion configurations consist
of ${n}$--molecules with fixed $n$  are connected regions
whose size decreases as $n$ increases. There is a relatively
large domain $D_{1}$ corresponding to neutral atoms
($\rho_{e}=\rho_{i}=\rho$) with $\rho$ ranging from some
$\rho_{min}(U)$ to $1-\rho_{min}(U)$,
a smaller domain $D_{2}$ corresponding to $2$--molecules with
$\rho_{i}=2\rho_{e}$, and consecutively domains $D_{n}$ corresponding
to $n$--molecules, where $n=2, 3, \ldots, N(U)$,
with $\rho_{i}=n\rho_{e}$ (i.e. in all the domains $D_{n}$ there is
one electron per molecule).
In each domain $D_{m}, m=1,\ldots, N(U)$ the ion density satisfies
the inequality
$\rho_{min}^{(m)}(U) \leq \rho_{i} \leq  \rho_{max}^{(m)}(U)$.
Furthermore the minimal value
$\rho_{min}^{(m)}(U)$ is zero for all $m$ between some $n(U)$ and
$N(U)$ (in the cases considered we had $n(U)=N(U)-1$) and for
those $m$ the boundary between $D_{m}$ and the vacuum
($\rho_{e}=\rho_{i}=0$, i.e. $\mu_{e}\leq -2$) is linear.
Between two domains $D_{n}$ and $D_{n+1}$ the periodic
configurations are such that their unit cells contain $n$ and
$(n+1)$--molecules only. Between $D_{n}$ and the vacuum the unit
cells consist of molecules and a sequence of empty sites while between
$D_{n}$ and the full configuration the unit cell contains atoms and
a sequence where all sites are occupied.

We note also that all the domains $D_{\rho_{e}}$
(with a given electron density), that appear in the restricted phase
diagrams, follow the Aubry sequence (Farey's tree, Fig. 10) \cite{aub}
with respect to $\rho_{e}$ and exhibit a devil's staircase structure
with respect to their size.
\begin{figure}
\vspace{9truecm}
\caption{Farey tree (or Aubry sequence) for electrons.}
\end{figure}

We therefore arrive at the conjecture that to all electron densities
between zero and one there correspond, according to the Aubry
sequence, the domains $D_{\rho_{e}}$ which present a fractal
structure known as the devil's staircase \cite{brubak}.

Similar observation can be made with respect to the structure of the
domains $D_{n}$ of $n$--molecules. The ion densities in these
domains appear to follow the "$n$--molecule Aubry sequence", that is
similar to the Aubry sequence, but starts with $0/n$ and $n/n$
and one has to consider rationals $p/q$ with $p=0 \;mod\; q$
(see Fig. 10).
This property can be understood as resulting from an effective
two--body interaction between $n$--molecules , that is convex and
decreasing (as was the case for the effective two--body interaction
between the atoms for large $U$). However, because of the effective
$k$--body interaction, we do not have the exact devil's staircase of
Burkov \cite{aub}. In the neutral case this effect was already
observed in \cite{llj}, where the periodic g.s.c. up to period 6 and
for large $U$ were studied.
To check the above conjecture we have taken into account some periodic
configurations with period larger than 10. The results are shown in
Fig. 11.
\begin{figure}
\vspace{18truecm}
\caption{Test of the conjecture regarding the Aubry sequence
and the devil's staircase.
(a) $U=0.6$,
(b) $U=0.6$,
(c) $U=0.2$,
(d) $\mu_{e}$--extents of the
domains contained in
$ D_{1} $. }
\end{figure}

An inspection of the phase diagrams (and Fig. 12) yields a more
detailed picture of the neutral case discussed within the canonical
formalism. For $U<2/\sqrt{3}$ the domain corresponding to
non--trivial
neutral configurations ($\rho_{e}=\rho_{i}$) consists of the curve
$AB$, part of the boundary between the empty
configuration and the periodic configurations, and the domain $D_{1}$
ending at $B$. Along the curve $AB$ the density
increases monotonically from zero to $\rho_{min}(U)$. At each point
on these curves the g.s.c.
is a mixture $[p/q]_{n}\&-$ or $[p/q]_{n}\&[p'/q']_{n+1}\&-$.
Furthermore at each point the density is different (because of the
weight of the vacuum, not because of the periodic g.s.c.).
In the domain $D_{1}$ the density is piecewise constant, increasing
from $\rho_{min}(U)$ to $1-\rho_{min}(U)$, with a devil's staircase
structure. The results for $\rho_{e}=\rho_{i}$ and $U=0.6, 0.4, 0.2$
have been reported in Fig. 5.
\begin{figure}
\vspace{18truecm}
\caption{The region of the $(\mu_{e}, \mu_{i})$ plane where
the system is neutral ($\rho_{e}=\rho_{i}$). In the hatched area the
g.s.c. are periodic while at the curve $AB$ the g.s.c.
are mixtures of periodic configurations with the empty configuration.}
\end{figure}

We conclude this section with an argument leading to an
equation for $\mu_{e}^{*}$, the $\mu_{e}$--coordinate of the point,
where the coexistence line of the empty and full configurations joins
the domain $D$. The reduced phase diagrams show that
for $\mu_{e}\in [ \mu_{e}^{*},-2]$
the boundary of $D$ with the vacuum ($\rho_{e}=\rho_{i}=0$)  is
piecewise linear and corresponds to the configurations of the form
$[n/\infty]_{n}$. On the other hand our considerations in
the canonical formalism have shown that for a finite number of ions
the g.s.c. of a neutral system consists of $n$--molecules
(where $n$ is U--dependent) with one bound electron
and $n-1$ diffusion electrons. For $\mu_{e}<-2$
there is no diffusion electrons and we then have only $n$--molecules
with one electron. The free energy per ion is thus

\be
\label{fen_nm}
f(H_{n};\mu_{e},\mu_{i})= {E^{(n)}_{0} - \mu_{e}\over
n} - \mu_{i} \ .
\ee
where $E_{0}^{(n)}$ is the energy of the $n$--molecule ground state.
On the other hand the free energy of the vacuum is zero. Therefore
the boundary of the vacuum domain (for $\mu_{e} \leq-2$) is given by the
concave envelope of the functions of $\mu_{e}$

\be
\label{ff=0}
f(+;\mu_{e},\mu_{i}) = 0 , \qquad
f(H_{n};\mu_{e},\mu_{i}) = 0, \,i.e. \qquad
\mu_{i}^{n}=\left( E_{0}^{n}-\mu_{e} \right)/n.
\ee
For $\mu_{e} =-2$ the value $n(U)$ that corresponds to the minimum of the
sequence $(E_{0}^{(n)}+2)/n$ is equal to the number of ions in
the molecule at zero density (Table \ref{b}). Since the function
given by $f(H_{n};\mu_{e},\mu_{i}) = 0 $ has the slope $-1/n$,
it is sufficient to construct the concave envelope of the functions
given by eq.(\ref{ff=0}) with $n\geq n(U)$.
These functions are plotted in Fig. 13.
\begin{figure}
\vspace{9truecm}
\caption{Graphs of the functions defined by eq. (26).}
\end{figure}
In particular it appears
that for $U\geq 2.65$ the value of $\mu_{e}^{*}$ is determined
by $f(H_{1};\mu_{e},\mu_{i}) = 0 $, i.e.

\be
\label{f1m=0}
\mu_{i} = - \sqrt{4 + U^{2} } - \mu_{e},
\ee
which together with eq.(\ref{d+d-}) gives the following equation for
$\mu_{e}^{*}$:

\be
\label{b_eq}
\begin{array}{c}
\displaystyle{
\sqrt{4 + U^{2}} = U + {1\over \pi}
\left( \sqrt{4 - {\tilde{\mu}}^{2}_{e} }
- \tilde{\mu}_{e}  arc \  \cos \left(
\tilde{\mu}_{e}/2 \right) \right) \ , }
\cr\cr
\displaystyle{
\tilde{\mu}_{e} = \mu^{*}_{e} + U \, ,
\qquad \left( U \geq 2.65 \right) }.
\end{array}
\ee
and the range of densities corresponding to the segregated
configuration:

\be
\label{rhos_seg2}
\rho_{e} \leq \rho_{i} {1 \over \pi}
arc \  \cos \left( - \tilde{\mu}_{e}/2 \right).
\ee
This curve is plotted in Fig. 14 together with the upper and lower
bounds derived by Brandt \cite{bra}. To the precision of Fig. 14
the curve obtained by Freericks and Falicov \cite{frefal} almost
coincides with our curve and only a few points have been reported.
\begin{figure}
\vspace{9truecm}
\caption{Canonical phase diagram. For $U>2/ \protect\sqrt{3}$ and
$\rho_{e}=\rho_{i}$ the g.s.c. is the atomic most homogeneous
configuration. The boundary of the segregated configuration, shown as
the continuous line, was computed from eq. (29).
The dashed curves represent the bounds obtained in [13]
while the crosses represent  numerical results of [6].}
\end{figure}

Finally we add a remark concerning the canonical phase diagrams
obtained by Freericks and Falicov \cite{frefal}.
Let us take $U=0.6$. For the densities
($\rho_{e}=1/3$, $\rho_{i}= 1/2)$ the unit cell of the g.s.c.
in \cite{frefal}
consists of three occupied sites followed by three empty sites;
the energy density of the corresponding g.s.c. is equal to
$-0.667401$.
We found that the even mixture $[1/3]\&[2/3]$ has the energy density
$-0.668167$.
Similarly, for the densities ($\rho_{e}=1/2$, $\rho_{i}= 1/3)$)
the unit cell of the g.s.c. in \cite{frefal}
has the length 6 and consists of two occupied sites separated by one
empty sites, the other sites being empty.
The energy density of the corresponding g.s.c. is equal to
$-0.753835$.
We found that the  mixture $[1/2]\& -$ has the energy density
$-0.754651$.

In conclusion, to be physically meaningful, an incoherent phase
diagram (in the plane $(\rho_{e}, U)$ and for fixed $\rho_{i}$ )
must take into account mixtures of configurations with different ion
densities. The same remark, as we have already seen, applies to the
neutral case.

\section{Conclusions}

Our  analysis  of the zero temperature restricted phase diagrams of
the spinless Falicov--Kimball model has revealed several new and
surprising effects.

In the case of \underline{neutral systems} we have seen that for
$U>2/\sqrt{3}$ the ground state configuration is, as expected,
atomic most homogeneous and the system is an insulator.
However for $U<2/\sqrt{3}$ this remains
true only for densities $\rho$ which are not too small or not too
large. Indeed there is a critical density $\rho_{c}(U)$ such that
for $\rho<\rho_{c}(U)$ (or $\rho>1- \rho_{c}(U)$) the ground state
presents entirely different properties. We observed three new
phenomena:

--{\em Phase separation}; the ground state configuration is a mixture
of a periodic configuration and the empty configuration, i.e.
all ions are distributed periodically in one semi--infinite
half--space while the other half--space is free of ions.

--{\em Formation of $n$--molecules}; the ions form clusters where
$n$ consecutive lattice sites are occupied and these clusters are
most homogeneously distributed over the occupied half--space.

--{\em Changes in conducting properties}; the Fermi level is located
in the band
of the empty configuration and therefore an extra electron can be added
at no cost in the energy.

As the density $\rho$  decreases the following infinite sequence of
phase diagram transformations is observed.
In the first stage the weight
of the half--space with no ions increases while the unit cell in the
half--space occupied by the ions remains unchanged.
In the second stage, at a sufficiently small density the distances
between the molecules jump to a larger value (modification of the
unit cell but not of the size of the molecules).
Then the two stages repeat themselves until the value of the density
is reached, where a transition from $n$--molecules to
$(n+1)$--molecules occurs. After this transition the first two stages
repeat themselves again, but now with  ion configurations containing
$(n+1)$--molecules, and so on. For a given potential $U$ there is a
largest admissible size $N(U)$ of the molecules that can be formed.
After the $N(U)$--molecules have been formed the remainder of the
sequence consists of the first two stages repeating themselves
indefinitely.

These phenomena were interpreted as consequences of the fact that
for small $U$ the $n$--molecules  have only one bound electron,
the other $n-1$ electrons being diffusive. The g.s.c. appears then
to result from the competition between a strictly convex effective
repulsion between the molecules and the pressure exerted by the
"free" electrons.

On the other hand, for any density $\rho$,
$\rho_{c}(U)< \rho <1-\rho_{c}(U)$, the ground state configuration
is periodic most homogeneous and the system is an insulator.
Furthermore in the plane of the chemical potentials the density in
the neutral domain appears to follow the Aubry sequence and the sizes
of phase domains show the devil's staircase structure.

We pass now to the  \underline{non--neutral systems}.
In this case we have found that for $\rho_{e} < \rho_{i} b(U)$,
where $b(U)$ is given by (\ref{rhos_seg2})
the g.s.c. is the segregated configuration and the system is a
conductor. The form of $b(U)$ has been checked to be in agreement
with our numerical results, the numerical results given in
\cite{frefal} and the bounds given in \cite{bra}.

For densities $\rho_{e}$ and $ \rho_{i}$ such that
$\rho_{i} b(U) < \rho_{e} < 1/2$, the  set of periodic g.s.c.
decomposes mostly into sequences for which $\rho_{i}=n \rho_{e},
1\leq n \leq N(U)$. Given such a sequence, then for any $\rho_{i}$
between $\rho_{min}^{n}(U)$ and $\rho_{max}^{n}(U)$, the ground state
consists of $n$--molecules which are most homogeneously distributed
over the lattice, with one electron per molecule.

For general values of $(\rho_{e}, \rho_{i})$ we came to the
conclusion that for any electron density $\rho_{e}=p/q$, with $p$
relatively prime to $q$, there are periodic ground states with
ion densities $\rho_{i}=\tilde{p}/q$, where $\tilde{p}$ is any
integer satisfying $p'\leq \tilde{p} \leq p''$. If $\rho_{i}$ is
such that $\tilde{p}/q < \rho_{i} < (\tilde{p}+1)/q$,
the ground state configuration is a mixture of the two phases with
density $\tilde{p}/q$ and $(\tilde{p}+1)/q$
(for $\tilde{p}=p''$ it is a mixture of the phase $p''/q$ and the full
configuration); for $\rho_{i} < p'/q$ it is a mixture of the phase
$p'/q$ and the empty configuration. In any case
if the  ground state is periodic, then the system is an insulator.

The restricted phase diagram presents the following general
properties. The non-trivial part of the $(\mu_{e}, \mu_{i})$ plane,
i.e.$-(U+2)<\mu_{e}<2$, decomposes into three connected domains
$D_{-}$, $D_{+}$ and $D$, corresponding respectively to the empty,
full and periodic translationally non--invariant configurations.
The domains $D_{-}$ and $D_{+}$ are separated by domain $D$, contained
in the vertical stripe $[\mu_{e}^{*}, -(U+\mu_{e}^{*})]$, where
$\mu_{e}^{*}=\mu_{e}^{*}(U)<min\{-2, -U/2 \}$, and by two curves
which correspond to the segregated phase (see Fig. 12).
There are two partitions of $D$ into connected subdomains that are of
importance in the analysis of the restricted phase diagrams.
In the first partition there appear subdomains
$D_{n}$, where the ions in the g.s.c. form  $n$--molecules with one
electron per molecule ($\rho_{i}=n\rho_{e}$). In each domain $D_{n}$
the ion density follows the Farey tree order (Aubry sequence)
and the domains of distinct periodic configurations exhibit the
devil's staircase structure.
The second partition consists of the subdomains $D_{\rho_{e}}$, where
the electron density has a definite rational value $\rho_{e}=p/q$.
These domains form "curved stripes" going across $D$ from the boundary
with $D_{-}$ to the boundary with $D_{+}$. The ion densities in these
domains constitute the sequence $p'/q, (p'+1)/q, \ldots, p''/q$,
for some $p'<p''$ that depend on $\rho_{e}$ \vbox{and $U$.}

Finally, we have seen that theorem 1 of \cite{frefal} is partially
incorrect and at the same time seems to be more general.
We conjecture that this theorem should be valid for all $U$ in the
form: Given $\rho_{e}=p/q$, with $p$ relatively prime to $q$, and
$\rho_{i}=\tilde{p}/q$, then if the ground state is periodic, it is
given by the Freericks--Falicov method.

\section*{Acknowledgments}
One of the authors (J.J.) acknowledges the financial support of the
Committee for Scientific Research (Poland) under grant 2 P302 147 04
and the hospitality of the
Institut de Physique Th\'{e}orique of the Ecole Polytechnique
F\'{e}d\'{e}rale de Lausanne, where this research was conducted.


\newpage

\begin{thebibliography}{99}

\bibitem[1]{falkim}
L.M. Falicov and J.C. Kimball,
Phys.\ Rev.\ Lett.{\bf 22}, 997(1969).

\bibitem[2]{khom}
D.I. Khomskii, in {\it Quantum theory of solids}.
ed. I.\ M.\ Lifshits, Mir, Moscow 1982.

\bibitem[3]{kenlieb}
T. Kennedy and E.H. Lieb, Physica \ A {\bf 138},
320 (1986).

\bibitem[4]{brasch}
U. Brandt and R. Schmidt, Z.\ Phys. \ B {\bf 63}, 45 (1986).

\bibitem[5]{jll}
J. J\c{e}drzejewski, J. Lach and R. {\L}y\.{z}wa,
Physica A {\bf 154}, 529 (1989)

\bibitem[6]{frefal}
J.K. Freericks and L.M. Falicov,
Phys.\ Rev. \ B {\bf 41}, 2163 (1990).

\bibitem[7]{brasch2}
U. Brandt and R. Schmidt, Z.\ Phys. \ B {\bf 67}, 43 (1987).

\bibitem[8]{gijl}
Ch. Gruber, J. Iwa\'{n}ski, J. J\c{e}drzejewski
and P. Lemberger, Phys.\ Rev.\ B\ {\bf 41}, 2198 (1990).

\bibitem[9]{lebmac}
J.L. Lebowitz and N. Macris, Ecole Polytechnique Federale
de Lausanne preprint.

\bibitem[10]{mm}
A. Messager and S. Miracle-Sole, Marseille University preprint.

\bibitem[11]{gru}
Ch. Gruber, Helvetica Physica Acta {\bf 64}, 668 (1991).

\bibitem[12]{lem}
P. Lemberger, J.\  Phys. \ A {\bf 25}, 715 (1992).

\bibitem[13]{bra}
U. Brandt, J.\ Low \ Temp.\ Physics \ {\bf 84}, 477 (1991).

\bibitem[14]{barsub}
M. Barma and V. Subrahmanyam,
Phase \ Trans.\ B \ {\bf 16}, 303 (1989).

\bibitem[15]{gjl}
Ch. Gruber, J. J\c{e}drzejewski and P. Lemberger,
J. \ Stat. \ Phys. \ {\bf 66}, 913 (1992).

\bibitem[16]{glm}
Ch. Gruber, J.L. Lebowitz and N. Macris,
Europhys. Lett. {\bf 21}, 389 (1993);
Phys.\ Rev.\ B \ {\bf 48}, 4312 (1993).

\bibitem[17]{ken}
T. Kennedy, University of Arizona preprint.

\bibitem[18]{llj}
J. Lach, R. {\L}y\.{z}wa and J. J\c{e}drzejewski,
Phys.\ Rev.\ B \ (1993).;
Acta Phys.\ Polonica \ A \ {\bf 84}, 327 (1993).

\bibitem[19]{lyz}
R. {\L}y\.{z}wa, Phys.\ Lett.\ A \ {\bf 164}, 323 (1992);
Physica \ A \ {\bf 192}, 231(1993).

\bibitem[20]{hub}
J. Hubbard, Phys.\ Rev.\ B\ {\bf 17}, 494 (1978).

\bibitem[21]{lieb} E.H. Lieb, private communication

\bibitem[22]{aub}
S. Aubry, J.\ Physique\ -- Letters \ {\bf 44}, L--247 (1983).

\bibitem[23]{brubak}
R. Bruinsma and P. Bak, Phys.\ Rev.\ B\ {\bf 27}, 5824 (1983).

\end{thebibliography}
\end{document}